# Photon Emission from a Parton Gas at Chemical Non-Equilibrium *

Christoph T. Traxler and Markus H. Thoma

*Institut für Theoretische Physik, Universität Giessen,*

*35392 Giessen, Germany*[†]

## Abstract

We compute the hard photon production rate of a chemically non-equilibrated quark-gluon plasma. We assume that the plasma is already thermally equilibrated, i. e. describable by a temperature, but with a phase-space distribution that deviates from the Fermi/Bose distribution by a time dependent factor (fugacity). The photon spectrum is obtained by integrating the photon rate over the space-time evolution of the quark-gluon plasma. Some consequences for ultrarelativistic heavy ion collisions are discussed.

## I. INTRODUCTION

Hard photons are a promising probe for the fireball created in ultrarelativistic heavy ion collisions [1]. As they leave the fireball without further interaction [2] they are probing the various stages of the collision directly. Concerning the quark-gluon plasma (QGP) phase, photon emission has been considered so far mostly at thermal and chemical equilibrium [3–5]. Here we will investigate the photon production from a chemically non-equilibrated QGP, starting from a simple model for the chemical equilibration of a parton gas in ultrarelativistic heavy collisions [6]. This model predicts that the QGP, possibly formed at RHIC and LHC, is always far from chemical equilibrium, showing a strong undersaturation of the phase space in particular for quarks, which are the source of photon emission. In the next section we present a short review of the parton chemistry model, before we calculate the hard photon production rate from a chemically non-equilibrated parton gas in section 3. In section 4 we show photon spectra, obtained by integrating the photon rate over the space-time evolution of the fireball, and discuss some consequences for RHIC and LHC. Similar investigations have been performed in ref. [7–9], to which we compare our results.

---

*supported by BMFT and GSI Darmstadt

[†]e-mail: chris.traxler@uni-giessen.de, thoma@theorie.physik.uni-giessen.de

## II. PARTON CHEMISTRY

The chemical equilibration of quarks and gluons has been described by means of rate equations for the quark and gluon phase space density after thermal equilibrium set in [6]. Here we speak of thermal equilibrium as soon as the momentum distribution of the partons becomes exponential and isotropic. According to the event generator HIJING [10] the primary hard parton collisions result already in an exponential $p_\perp$-distribution [11]. Subsequent longitudinal expansion leads to an isotropic momentum distribution in the central slice ($\Delta z = 0.5$ fm) at a time $\tau_{iso} = 0.3$ fm/c at RHIC and $\tau_{iso} = 0.2$ fm/c at LHC after the primary collisions [6] corresponding to $\tau_0 = 0.5 - 0.7$ fm/c after the maximum overlap of the nuclei [11,12].

For times $\tau > \tau_{iso}$ we assume that the distribution functions can be approximated by

$$n_{F,B}(E) = \lambda_{q,g}(\tau) \frac{1}{e^{E/T(\tau)} \pm 1}, \qquad (1)$$

i.e. by equilibrium Fermi/Bose distribution functions with a time dependent temperature multiplied by a time dependent factor $\lambda_{q,g}$ which describes the deviation from chemical equilibrium. This factor, called fugacity, takes account of the undersaturation of the parton phase space density, i.e. $0 \leq \lambda_{q,g} \leq 1$.

To lowest order perturbative QCD the phase space will be populated by the reactions

$$gg \leftrightarrow ggg, \qquad gg \leftrightarrow q\bar{q}. \tag{2}$$

The time evolution of the quark and gluon densities, which are proportional to the fugacities, can de determined by rate equations, where the equilibration rates entering these equations follow from the cross section of the above reactions. The cross sections are calculated from the lowest order matrix elements, where the thermal gluon and quark mass, also depending on the fugacities, serve as infrared cut-offs:

$$m_g^2 = \lambda_g \left(1 + \frac{N_f}{6}\right) \frac{g^2 T^2}{3}, \tag{3}$$

$$m_q^2 = \left(\lambda_g + \frac{\lambda_q}{2}\right) \frac{g^2 T^2}{9}, \tag{4}$$

where $N_f$ denotes the number of active flavors in the parton gas, for which $N_f = 2.5$ was chosen. (In chemical equilibrium the thermal masses follow from the zero momentum limit of the gluon and quark self energies in the high temperature approximation [13] leading to

$$m_g^2 = \frac{2g^2}{\pi^2} \left(1 + \frac{N_f}{6}\right) \int_0^\infty dk\, k\, n_B^{eq}(k),$$

$$m_q^2 = \frac{2g^2}{3\pi^2} \int_0^\infty dk\, k\, \left[n_B^{eq}(k) + n_F^{eq}(k)\right].$$

In the chemically non-equilibrated parton gas the equilibrium distribution functions are replaced by (1) leading to (3) and (4).) In addition, gluon radiation in the first reaction is assumed to be suppressed if the mean free path of the gluons in the parton gas is smaller than the formation length of the emitted gluon. In this way the so called Landau-Pomeranchuk effect is taken into account phenomenologically. The results for the equilibration rates can be found in ref. [6].

Together with the equation describing energy conservation in the case of a purely longitudinal expansion [6] the rate equations determine the evolution of the quark and gluon fugacities $\lambda_{q,g}(\tau)$ and of the temperature $T(\tau)$. These equations are solved numerically together with the following initial conditions at $\tau_{iso}$ resulting from HIJING: $\lambda_g^0 = 0.09$, $\lambda_q^0 = 0.02$, and $T_0 = 570$ MeV for RHIC and $\lambda_g^0 = 0.14$, $\lambda_q^0 = 0.03$, and $T_0 = 830$ MeV for LHC, respectively. The results are shown in fig. 1. (Recent investigations [12] with slightly changed initial conditions and a somewhat larger rate for the gluon production gave similar results). The temperature drops faster than in the Bjorken scenario [14] ($T^3\tau$=const), since energy is consumed by parton production. The main result of these investigations is a clear deviation from chemical equilibrium (undersaturation) at RHIC as well as LHC, especially for quarks.

The following problems and criticisms associated with this model and its results should be mentioned:

1. The initial fugacities from HIJING are very small in contrast to the one following from the Parton Cascade Model [15], which gives $\lambda_g^0 \simeq 1$ and $\lambda_q^0 \simeq 0.7$. The reason for this essential difference is not clear yet.

2. The introduction of time dependent distribution functions such as (1) for describing a non-equilibrium situation contradicts a perturbative expansion in the real time formalism by leading to singularities in loop diagrams [16]. However, here we only want to employ the distributions (1) as a phenomenological ansatz, and we do not consider loop diagrams.

3. Uncertainties in the equilibration rates have a significant influence on the evolution of the fugacities. In particular, a larger gluon production rate could be obtained if the Landau-Pomeranchuk effect in the gluon bremsstrahlung is treated by taking into account the rescattering of the radiated gluon. Then gluons with larger formation times can also be emitted [17]. Furthermore, the process $gg \leftrightarrow ng$, $n > 3$, can be included leading to gluon production rates about twice as large [18].

Despite all these uncertainties, we think it worthwhile to study the consequences of this parton chemistry scenario, as it has been done in the case of charm production [12]. The other way round, by comparing the particle production (photons, dileptons, charm), predicted by this scenario, with experimental data, one might be able to extract information on the equilibration in ultrarelativistic heavy ion collisions.

## III. PHOTON PRODUCTION RATE

In thermal and chemical equilibrium, the production rate of hard photons with energy $E \gg T$ can be computed using the Braaten-Yuan prescription [19]. This results in a decomposition of the rate into a soft part, which is treated using a resummed quark propagator according to the Braaten-Pisarski method [20], and a hard part containing only bare propagators and vertices. In the soft part the resummed quark propagator takes care of medium effects in the QGP; e.g., it contains the thermal quark mass $m_q^2 = g^2 T^2/6$, which serves as an infrared cut-off in the case of a vanishing bare quark mass. The hard part follows from the momentum integration over the matrix elements that lead to photon emission in lowest order (quark pair annihilation, Compton scattering with an initial gluon) multiplied by the distribution functions of the incoming and outgoing partons [3]. A separation parameter $k_c$ is introduced, which allows to distinguish between soft and hard momenta of the intermediate quark. Assuming the weak coupling limit and $gT \ll k_c \ll T$, the final result is independent of the separation scale $k_c$. This procedure has been demonstrated in refs. [3,4] using Boltzmann distribution functions for the incoming particles, and in ref. [5] using full Fermi/Bose distribution functions with a nonzero quark chemical potential $\mu$. Unfortunately, the Braaten-Pisarski method is based on the principle of detailed balance, which holds only in full equilibrium, and thus is not applicable for the chemical non-equilibrium stage.

Since there exists so far no consistent method for treating medium effects at non-equilibrium, we propose the following procedure. We only consider the hard part of the photon rate and replace the cut-off $k_c^2$ by $2m_q^2$ as suggested by the equilibrium result [3]. For the thermal quark mass we adopt the formula (4) containing the fugacities, thus taking into account non-equilibrium effects. This approximation is in line with the estimates of the equilibration rates calculated in ref. [6]. In addition, we assume $\mu = 0$ since the photon rate is not sensitive to a non-zero quark chemical potential [5]. Furthermore, the fugacities show up in the distribution functions of the incoming and outgoing partons, for which we adopt the non-equilibrium distributions (1), under the momentum integral defining the photon rate.

We now compute the hard photon production rate, starting from the following equation for each contribution (annihilation, Compton) of the photon rate [5] (summed over the photon polarisations):

$$2E\frac{dn}{d^3p\,d^4x} = \frac{1}{8(2\pi)^7 E}\int_{2k_c^2}^{\infty} ds \int_{-s+k_c^2}^{-k_c^2} dt \sum |\mathcal{M}|^2 \int_{\mathrm{IR}^2} dE_1\, dE_2 \frac{\Theta(P(E_1,E_2))n_1 n_2(1\pm n_3)}{\sqrt{P(E_1,E_2)}}, \quad (5)$$

where $\sum |\mathcal{M}|^2$ denotes the square of the matrix element for the annihilation or Compton process summed over the initial and final parton states and $n_i$ the parton distributions, where the plus sign in front of $n_3$ corresponds to the annihilation process and the minus sign to Compton scattering. The polynomial $P$ is given by $P(E_1, E_2) = -(tE_1 + (s+t)E_2)^2 + 2Es((s+t)E_2 - tE_1) - s^2 E^2 + s^2 t + st^2$ with the Mandelstam variables $s,t$, and $\Theta$ is the step function. $E$ and $\boldsymbol{p}$ are the outgoing photon energy and momentum, related by $E = p \equiv |\boldsymbol{p}|$, since the photon is on-shell. Finally, $dn$ is the number of photons that will come out of the plasma cell $d^3x$ during $dt$ with a momentum in $d^3p$ around $\boldsymbol{p}$. Equation (5) is already written in a Lorentz invariant fashion (for boosts along the beam direction) and can be related to the rapidity via $E/d^3p = dp_x dp_y dy = d^2 p_T dy$.

Let us now introduce the fugacity factors in (5) by the replacement

$$n_1 n_2 (1 \pm n_3) \;\mapsto\; \lambda_1 n_1 \lambda_2 n_2 (1 \pm \lambda_3 n_3) \quad.$$

We decompose this product as follows:

$$\lambda_1 n_1 \lambda_2 n_2 (1 \pm \lambda_3 n_3) = \lambda_1 \lambda_2 \lambda_3 n_1 n_2 (1 \pm n_3) + \lambda_1 \lambda_2 (1 - \lambda_3) n_1 n_2 \quad. \quad (6)$$

Here, the first term leads to the equilibrium photon rate as computed in ref. [3] multiplied by $\lambda_q^2 \lambda_g$:

$$\left(2E\frac{dn}{d^3p\,d^4x}\right)_1 = \frac{5\alpha\alpha_s \lambda_q^2 \lambda_g}{9\pi^2} T^2 e^{-E/T} \left[\ln\left(\frac{4ET}{k_c^2}\right) - 1.42\right] \quad. \quad (7)$$

Here $k_c^2$ is the infrared cut-off of the hard contribution for which we will use $2m_q^2$ as discussed above. In order to find (7), one has to use Boltzmann distribution functions instead of the full quantum-mechanical distribution functions on the incoming legs. This underestimates the contribution coming from the Compton scattering process and overestimates the contribution of the quark-antiquark-annihilation process, but the two errors cancel up to a maximum error of about 10% in the end, as is shown by numerical analysis in ref. [5].

The second term of (6) results in the rate

$$\left(2E\frac{dn}{d^3p d^4x}\right)_2 = \sum_{\substack{ann\\comp}} \frac{\lambda_1 \lambda_2 (1-\lambda_3)}{8(2\pi)^7 E}$$

$$\times \int_{2k_c^2}^{\infty} ds \int_{-s+k_c^2}^{-k_c^2} dt \sum |\mathcal{M}|^2 \int_{E+s/4E}^{\infty} dE_+ \int_{IR} dE_2 \frac{\Theta(P(E_+,E_2)) n_1 n_2}{\sqrt{P(E_+,E_2)}} \quad . \tag{8}$$

We treat this term again in the Boltzmann approximation, using $n_1 n_2 = e^{-E_+/T}$. The distribution functions are constant in the innermost integral, and can be drawn out in front. Now all the integrals are elementary, and (8) reduces in a few steps to

$$\left(2E\frac{dn}{d^3p d^4x}\right)_2 = \frac{10\alpha\alpha_s}{9\pi^4} T^2 e^{-E/T}$$

$$\times \left\{ \lambda_q \lambda_g (1-\lambda_q) \left[ 1 + 2e^{-k_c^2/4ET} E_1\left(\frac{k_c^2}{4ET}\right) \right] \right.$$

$$\left. + \lambda_q \lambda_q (1-\lambda_g) \left[ -2 + 2e^{-k_c^2/4ET} E_1\left(\frac{k_c^2}{4ET}\right) \right] \right\} \quad . \tag{9}$$

For a consistent computation to order $\mathcal{O}(\alpha_s)$, we approximate

$$2e^{-k_c^2/4ET} E_1\left(\frac{k_c^2}{4ET}\right) \approx -2\gamma + 2\ln\left(\frac{4ET}{k_c^2}\right) + \mathcal{O}(\frac{k_c^2}{4ET})$$

and obtain

$$\left(2E\frac{dn}{d^3p d^4x}\right)_2 = \frac{10\alpha\alpha_s}{9\pi^4} T^2 e^{-E/T} \left\{ \lambda_q \lambda_g (1-\lambda_q) \left[ 1 - 2\gamma + 2\ln\left(\frac{4ET}{k_c^2}\right) \right] \right.$$

$$\left. + \lambda_q \lambda_q (1-\lambda_g) \left[ -2 - 2\gamma + 2\ln\left(\frac{4ET}{k_c^2}\right) \right] \right\} \quad . \tag{10}$$

The photon rate is the sum of (7) and (10) after insertion of $k_c^2 = 2m_q^2 = 0.22 g^2 T^2 (\lambda_g + \lambda_q/2)$.

Our result for the non-equilibrium photon rate differs from the one found in refs. [7,8], where the equilibrium rate was simply multiplied by the fugacities of the incoming partons and the equilibrium quark mass was used as infrared cut-off. Our formula also differs from the one of ref. [9], where more elaborate non-equilibrium distributions (Jüttner distributions) were assumed, which are, however, not in line with the parton chemistry model presented in ref. [6].

## IV. PHOTON SPECTRUM

In the simplest of all models of a central ultrarelativistic heavy ion collision, one imagines two flat nuclei which penetrate each other and fly apart afterwards, creating a longitudinally expanding, cylindrical quark-gluon plasma tube between them. A Minkowski diagram of this event is shown in fig.2.

For the total photon yield, we take the rate as given in the last section and the fugacities shown in fig.1, and integrate over the plasma space-time volume, using

$$\int_{plasma} d^4x = Q \int_{\tau_0}^{\tau_1} d\tau\, \tau \int_{-y_{nuc}}^{y_{nuc}} dy\ ,$$

where the times after the maximum overlap of the nuclei are $\tau_0 = 0.7$ fm/c and $\tau_1 = 4$ fm/c and the rapidity of the nuclei is $y_{nuc} = 6$ at RHIC and $\tau_0 = 0.5$ fm/c, $\tau_1 = 6.25$ fm/c, and $y_{nuc} = 8.8$ at LHC, respectively. $Q$ is the transverse cross section of the nuclei; for gold, $Q \approx 180 fm^2$. We obtain for the photon spectra

$$\left(\frac{2dn}{d^2p_\perp dy}\right)\bigg|_{y,p_\perp} = Q \int_{\tau_0}^{\tau_1} d\tau\, \tau \int_{-y_{nuc}}^{y_{nuc}} dy' \left(2E\frac{dn^{loc.rest}}{d^3p d^4x}\right)\bigg|_{\substack{E^{loc.rest} = p_\perp \cosh(y-y'),\\ T=T(\tau),\, \lambda=\lambda(\tau)}} . \quad (11)$$

(The two expressions in large round brackets are Lorentz scalars; we can Lorentz transform them by simply transforming their argument, i.e., the photon energy. The photon energy in the laboratory frame equals $p_\perp \cosh(y)$; the photon energy in the comoving frame of the plasma is then $p_\perp \cosh(y-y')$. The superscript *loc.rest* serves to remind us of this transformation.) Here we assumed that we can use the fugacities for times $\tau > \tau_0$ not only for the central region and neglected the transverse expansion of the fireball, which can be treated in a hydrodynamical model [21].

Fig.3 shows the individual contributions of different time intervals between $\tau_0$ and $\tau_1$ to the photon spectrum. Summing up these contributions gives rise to the concave shape of the photon spectrum in fig.3. Clearly the spectrum is dominated by early times corresponding to high temperatures [8], especially for large photon energies. Hence it is rather insensitive to the uncertainties of the equilibration rates. Compared to chemical equilibrium $\lambda_g = \lambda_q = 1$ at an initial temperature of $T_0 = 300$ MeV, as it was considered for SPS energies [21], the photon yield from the QGP is suppressed by about a factor of $10^{-2}$ (RHIC) to $10^{-1}$ (LHC). This would imply that photons from the QGP cannot be observed. Although the photon rate is enhanced by the high initial temperature $T_0 = 500 - 800$ MeV, this increase is overcompensated by the small fugacities of our model. Using the fugacities as plotted in ref. [12], the photon yield is the same as in fig.3 within a factor of 2.

Spectra for different rapidity regions are plotted in fig. 4. Within the limits given by the nuclear rapidities, the transverse momentum spectra do not depend on the photon rapidity. In other words, the photon rapidity distribution closely resembles the underlying quark rapidity distribution [22]. This is easily understood, as the rapidity distribution of massless isotropic radiation is *always* a bell-shaped curve with a width at half maximum (FWHM) $\Delta y \approx 1.6$. This is small compared with the nuclear rapidities ($y_{nuc} = 6 - 8.8$) at high energies. Therefore, if we Lorentz transform (shift) the rapidity distribution of the photons according to the velocity of the plasma cell they originate from, we get essentially the plateau-shaped distribution of the quarks, smeared out a little bit. This is schematically shown in fig. 5.

## V. CONCLUSIONS

We have studied the photon production from a chemically non-equilibrated parton gas presumably produced at RHIC and LHC. We have neglected the pre-thermal stage as well as the mixed and hadronic phase of the fireball. Using the undersaturated parton densities, following from rate equations of the parton chemistry based on initial conditions predicted by HIJING, the photon yield is suppressed by a factor $10^{-2}$ (RHIC) to $10^{-1}$ (LHC) compared to a fully equilibrated QGP at an initial temperature of $T_0 = 300$ MeV. This large difference relies on the fact that the HIJING model predicts a very small quark density (only 2 - 3% of the equilibrium value) at the time $\tau_0$, from which on the parton distributions look thermal, and that the photon emission is dominated by early times. This result suggests that the photon emission from the plasma phase is not observable. In contrast, the Parton Cascade Model predicts almost complete saturation of the parton phase space densities at the onset of thermal equilibrium with similar initial temperatures ($T_0 = 500$ - 800 MeV). Hence the photon yield from the plasma is enhanced by about a factor of $10^2$ (RHIC) to $10^3$ (LHC) compared with our estimate and might be visible, in particular for photon energies between 2 - 3 GeV [1]. Consequently, photon data at RHIC and LHC might allow a distinction between the initial condition predicted from HIJING and the one from the Parton Cascade Model.

# REFERENCES

[1] P. V. Ruuskanen, Nucl. Phys. A **544**, 169c (1992).

[2] M. H. Thoma, Phys. Rev. D **51**, 862 (1995).

[3] J. Kapusta, P. Lichard, and D. Seibert, Phys. Rev. D **44**, 1298 (1991).

[4] R. Baier, H. Nakkagawa, A. Niégawa, and K. Redlich, Z. Phys. C **53**, 433 (1992).

[5] C. T. Traxler, H. Vija, and M. H. Thoma, Phys. Lett. B **346**, 329 (1995).

[6] T. S. Biró, E. van Doorn, B. Müller, M. H. Thoma, and X.-N. Wang, Phys. Rev. C **48**, 1275 (1993).

[7] E. Shuryak and L. Xiong, Phys. Rev. Lett. **70**, 2241 (1993).

[8] B. Kämpfer and O. P. Pavlenko, Z. Phys. C **62**, 491 (1994).

[9] M. Strickland, Phys. Lett. B **331**, 245 (1994).

[10] X. N. Wang and M. Gyulassy, Phys. Rev. D **44**, 3501 (1991).

[11] K. J. Eskola and X. N. Wang, Phys. Rev. D **49**, 1284 (1994).

[12] P. Lévai, B. Müller, and X. N. Wang, LBL-preprint LBL-36594, 1994.

[13] V. V. Klimov, Zh. Eksp. Teor. Fiz. **82**, 336 (1982) [Sov. Phys. JETP **55**, 199 (1982)]; H. A. Weldon, Phys. Rev. D **26**, 1394 (1982); **26**, 2789 (1982).

[14] J. D. Bjorken, Phys. Rev. D **27**, 140 (1983).

[15] K. Geiger, CERN-preprint CERN-TH.7313/94, 1994 (submitted to Phys. Rep.).

[16] T. Altherr and D. Seibert, Phys. Lett. B **333**, 149 (1994).

[17] R. Baier, Yu. L. Dokshitzer, S. Peigné, and D. Schiff, Phys. Lett. B **345**, 277 (1995).

[18] L. Xiong and E. Shuryak, Phys. Rev. C **49**, 2203 (1994).

[19] E. Braaten and T. C. Yuan, Phys. Rev. Lett. **66**, 2183 (1991).

[20] E. Braaten and R. D. Pisarski, Nucl. Phys. **B337**, 569 (1990).

[21] N. Arbex, U. Ornik, M. Plümer, A. Timmermann, and R. M. Weiner, Phys. Lett. B **345**, 307 (1995).

[22] A. Dumitru, U. Katscher, J. A. Maruhn, H. Stöcker, W. Greiner, and D. H. Rischke, Frankfurt preprint UFTP-381/1995, 1995.

FIGURES

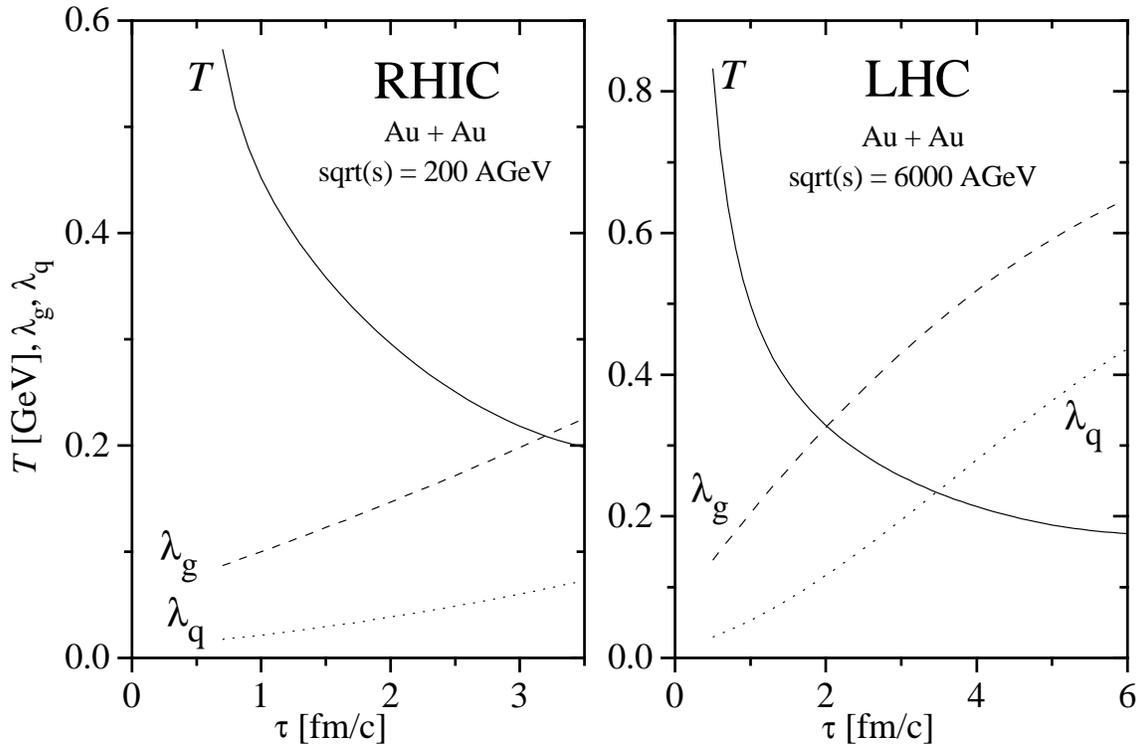

FIG. 1. Temperature, quark fugacity, and gluon fugacity for an Au + Au−collision in RHIC ($s^{1/2} = 200\ A \cdot GeV$) and LHC ($s^{1/2} = 6000\ A \cdot GeV$). The data was taken from [6].

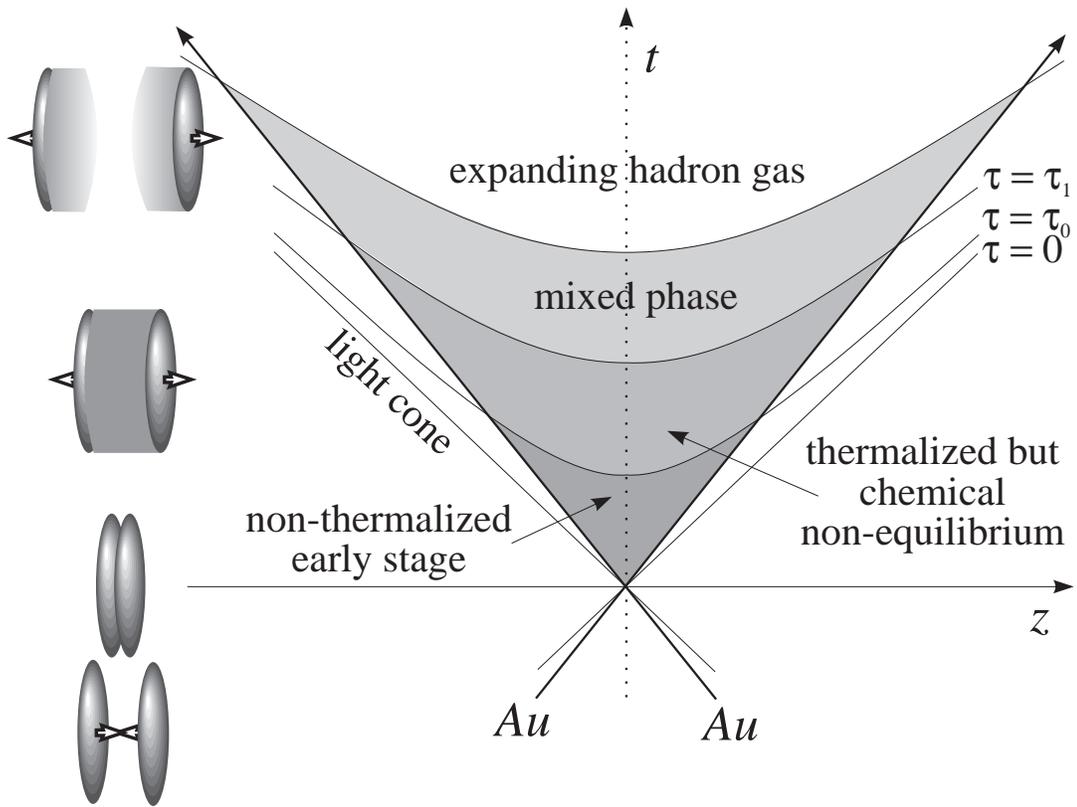

FIG. 2. *Minkowski diagram of a relativistic heavy ion collision (RHIC) creating a quark-gluon plasma. We compute the photon yield of the plasma between times $\tau_0$ and $\tau_1$.*

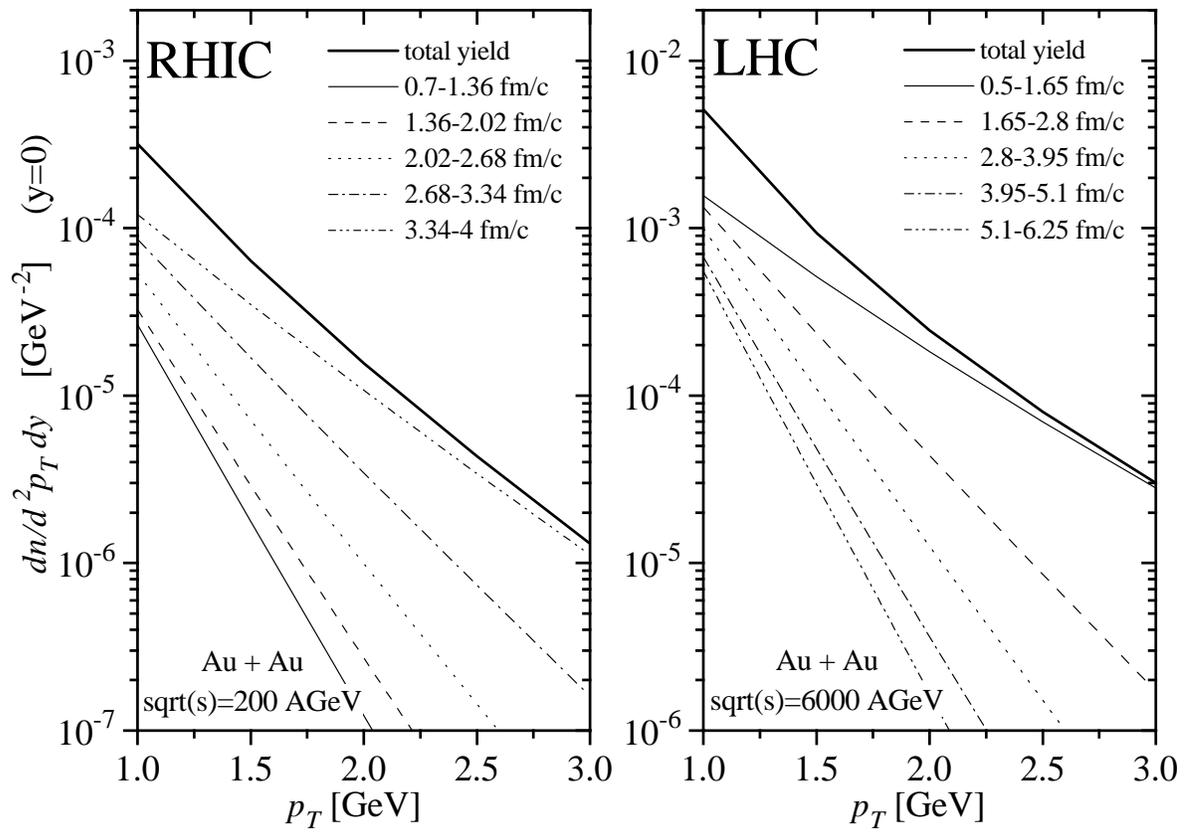

FIG. 3. *Midrapidity (y=0) photon yield from various stages (proper time slices) of the plasma evolution. The top curve is identical to the y=0 curve in the next figure.*

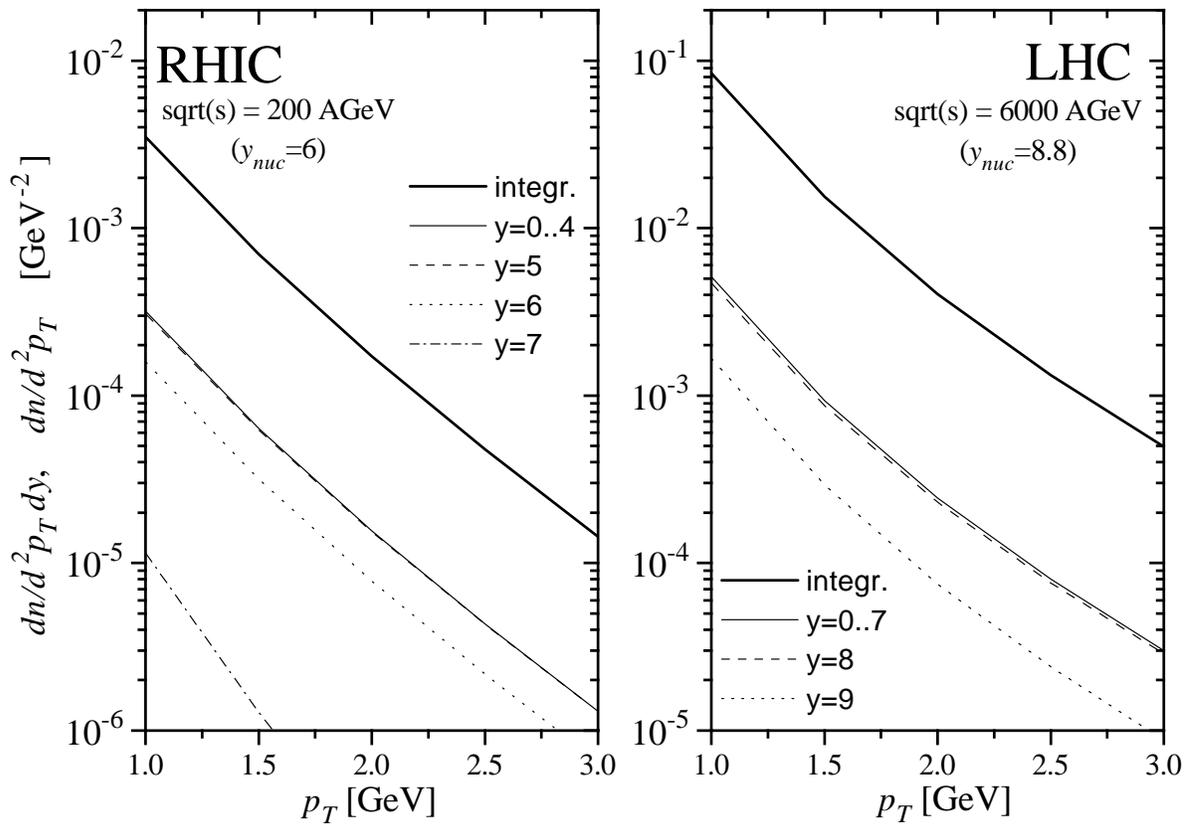

FIG. 4. *Photon spectra for various fixed rapidities.*

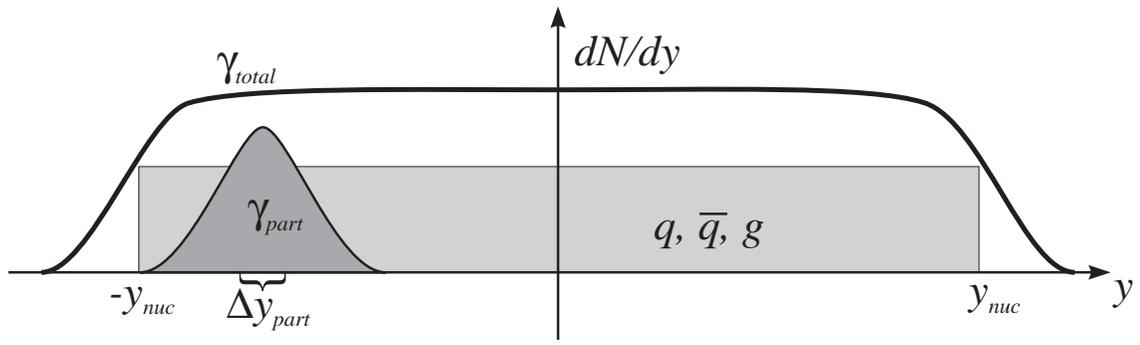

FIG. 5. *Rapidity distributions of photons, partons, and those photons originating from a parton cell within the rapidity interval $\Delta y_{part}$.*